# SD-EQR: A New Technique To Use QR Codes[TM] in Cryptography

Use of QR Codes[TM] In Data Hiding and Securing

Somdip Dey (Author)
Department of Computer Science
St. Xavier's College [Autonomous]
Kolkata, India.
Email: somdip007@hotmail.com

*Abstract*— **In this paper the author present a new technique of using QR Codes (commonly known as 'Quick Respond Codes') in the field of Cryptography. QR Codes are mainly used to convey or store messages because they have higher or large storage capacity than any other normal conventional 'barcodes'. In this paper the primary focus will be on storing messages in encrypted format with a password and send it to the required destination hiding in a QR Code, without being tracked or decrypted properly by any hacker or spyware. Since QR Codes have fast response time and have large storage capacity, QR Codes can be used perfectly to send encrypted data (messages) to the receiver. This method will be suitable in any business house, government sectors, communication network to send their encrypted messages faster to the destination. Or a person can even use this method to keep his important documents, like passport number, pan-card id, social security number, perfectly secured with him all the time, without the information getting leaked to outside world. The new method is achieved by entering the message along with a password. This password will generate a secret code, which will be added to each digit or alphabet in the numbers or text entered in the message (which is needed to be encrypted) and generate the first phase of encryption. That newly generated encrypted message will again be encrypted using various other methods to generate the final encrypted message.**

## A. Introduction

Data encryption is essential and a very important topic in modern day communication. It is very important to encrypt messages while sending them to a client from another client without being tracked or interpreted by a hacker. For example we can assume the situation where a bank manager is instructing his subordinates to credit an account, but in the mean while a hacker interpret the message and he uses the information to debit the account instead of crediting it. Again we can think of a situation, where a person wants to keep his passport information and other important documents safely secured with him all the time but he can not, may be because he is always exposed to outside intruders and threats. For this reason it is essential to use encryption to hide important data from the outside world and send it safely to the destination or keep it safe somewhere else. Encryption techniques are now a very important research field and, every now and then cryptography scientist are trying to come up with a good encryption technique (algorithm) so that no hacker / intruder can interpret the encrypted message. The modern day cryptographic methods are of two types (i) symmetric key cryptography, where the same key is used for encryption and for decryption purpose.

(ii) Public key cryptography, where we use one key for encryption and one key for decryption purpose. Symmetric key algorithms are well accepted in the modern communication network. The main advantage of symmetric key cryptography is that the key management is very simple. Only one key is used for both encryption as well as for decryption purpose. There are many methods of implementing symmetric key. In case of symmetric key method, the key should never be revealed / disclosed to the outside world other the user and should be kept secure. To deal with this problem we have introduced a new method of generating a code from the entered password, which will act as a key. In this present method the key generated from the password will act as first level of security of the encrypted message.

After doing the first level of encryption to the entered message using the symmetric key, many other several encryption techniques are used to encrypt the message further to increase the level of security, and this will make the job of an intruder or hacker nearly impossible to decrypt the encrypted message. At last the encrypted message will be converted to QR Code(s) and then send to the destination (client). Since maximum storage offered by QR Codes is 1,264 characters of ordinary / ASCII text and is only achieved in Version 40, error correction level H. For this reason if the size of the encrypted message is larger than 1,264 characters then many other QR Codes are generated containing the encrypted message after 1,264 characters and this method is continued until and unless the entire encrypted message is converted into QR Codes.

The different methods used to encrypt the message after the use of symmetric key for first level of encryption are:
(i) The encrypted message is then treated as a large string and the reverse of the string is generated.
(ii) The reverse encrypted string then extracted bit wise and XOR operation is performed with '1' (one) to those bits. This will generate another new encrypted message.

Then that encrypted message is converted into QR Code(s). Since QR Codes are good in hiding the message, that is why it can be used to hide the encrypted message in the QR Code.

And now to decrypt the message, first one has to know the key (password) or else the encrypted message will be shown and won't get the real message. After the key is entered to reveal the encrypted message the code will be generated and the encryption technique will be reverse processed to get back the original message.

B. ENCRYPTION (ENCRYPTING THE MESSAGE)

I. USE OF SYMMETRIC KEY

A. Generation of Code from Symmetric Key

We can choose a password of any size, but should consist of any ASCII character (0-255). Then from the entered password we need to find out the length of the password, which is denoted by 'plen'.

Now we will generate a number from the password by multiplying ($peln^2$) with each of the ASCII value of each letter of the word entered in the password and then adding the sum of all the values. Then to generate the code, each digit of the number generated is added with each other. The ultimate sum is the generated code.

Formula:
Let '$D_1D_2D_3D_4......D_{plen}$' be the password, where each $D_i$ ( i=1 to plen ) is a letter of the password string.
So the formula for generating the code:

$N = \lim i \to 1 \text{ to } plen \sum D_i \times (plen)^2$,
N=Number generated from the password

**Code** $= \lim i \to 1 \text{ to } len \sum n_i$,
N=$n_1n_2n_3n_4..........n_{(len)}$ and
$n_{(len)}$ = last digit of the number N
For Example:
Let the chosen password be 'Hello World'
Therefore, 'Hello World' is of length= 11 (including the blank space in between o and W of the string). So, plen =11.
Hence, from the above formula, the number generated is:

$72*(11)^2 + 101*(11)^2 + 108*(11)^2 + 108*(11)^2 + 111*(11)^2 + 32*(11)^2 + 87*(11)^2 + 111*(11)^2 + 114*(11)^2 + 108*(11)^2 + 100*(11)^2 = 127292$ ;
where the ASCII values of the letters are:
H=72; e=101; l=108; l=108; o=111; 'space'=32; W=87; o=111; r=114; l=108; d=100.
Therefore the required: Code= 1+2+7+2+9+2;
Code = 23.

B. Addition of Code with each Letter

The Code generated from the Symmetric Key (password) is now added to the ASCII value of each letter of the message to be encrypted. Thus the result is a new 'Value', generated due to the summation of the Code and each letter of the message.
After the generation of the 'Value', each 'Value' is converted to Unicode format by adding '&#' to the Value and the new encrypted message is formed, which is of Unicode format.

Formula:
Let '$L_1L_2L_3L_4.........L_n$' be the message, which is to be encrypted.
Therefore from the above theory,
Value = Code + ASCII($L_i$);
UniCode($L_i$) = '&#Value'; where i = 1 to n and $L_n$ = last letter of the last word of the message to be encrypted.
For Example:
Let the password be 'Hello World' as seen before. Therefore, the Code generated is equal to 23, i.e. Code= 23.
Let the message be 'I love you ÿþý'. Therefore, after adding Code with each of the ASCII value of each letter of the message, becomes:
Code + ASCII(I) = 23 + 73 = 96;
Code + ASCII('space') = 23 + 32 = 55;
Code + ASCII(l) = 23 + 108 = 131;
Code + ASCII(o) = 23 + 111 = 134;
Code + ASCII(v) = 23 +118 = 141;
Code + ASCII(e) = 23 + 101 = 124;
Code + ASCII('space') = 23 + 32 = 55;
Code + ASCII(y) = 23 + 121 = 144;
Code + ASCII(o) = 23 + 111 = 134;
Code + ASCII(u) = 23 + 117 = 140;
Code + ASCII('space') = 23 + 32 = 55;
Code + ASCII(ÿ) = 23 + 255 = 278;
Code + ASCII(þ) = 23 + 254 = 277;
Code + ASCII(ý) = 23 + 253 = 276;
Therefore the UniCode generated from the above Values -> ` 7 ƒ †  | 7  † Œ 7 Ė ĕ Ĕ (respectively) by adding '&#' in front of 'Value'.

Thus as a result, the original message 'I love you ÿþý' after encryption with the symmetric key use becomes:

`7ƒ†  |7    7ĖĕĔ  (in Unicode format)

II. REVERSING THE ENCRYPTED MESSAGE

Thus the first level of encryption is done by applying the symmetric key and the encrypted message is now again further encrypted so that it becomes very tough for the hacker or intruder to interpret the message.

So in this method the encrypted message is now treated as a string and the string is reversed. Let the encrypted string be in array format, where it starts from a[0] and ends at a[n-1], and n= total number of letters in the encrypted string. Thus the formula is the content of a[0] and a[n-1]are swapped, then the content of a[1] and a[n-2] are swapped and so on until
Function to implement reverse of a string:

```
function void reverse (char *s)
{
        int i; char temp;
        size_t  n= strlen(encrypted string);
        for (i=0; i <=n/2; ++1) {
                temp= a[i];
                a[i]= a[n-1-i];
                a[n-1-i]=temp;
                }
}
```

The above mentioned function can reverse the string totally, for example:
Let the encrypted message be:

`7ƒ†  |7      7ÈěĚ
Then, the reversed encrypted message is:

   ĚěĚ7     7|   † ƒ7`

### III.   XOR-ing The Encrypted Message

Since in this method of encryption Unicode format is used, so it is not wise to use the normal conventional way of representing the bits of the message as 8-bit binary data. The use of Unicode format make the scope of the encryption very broad, but at the same time it limits the use of normal bit wise encryption techniques on the message. Since Unicode characters are often larger than 8-bit binary values, so instead of 8-bit we will be using 16-bit binary values, i.e. the encrypted message will further be broken down in 16-bit binary values. After the letter are extracted as 16-bit binary numbers, they are XOR (exclusive OR) with '1'. And then the new 16-bit XOR-ed letters are put back in the message.
For example:
Ě = 0000000100010100 (16-bit binary format)
   Ě XOR (1) = 1111111011101011.
   Thus the new element becomes:  ♠ , which is ->
   ♠ = ( Ě XOR (1) ).
Therefore, the encrypted message ĚěĚ7    7|  † ƒ7`
after the application of XOR operation with '1' becomes:

౸▢ウケッ▢テウ ▢▢▢°

### C.  QR Code Generation and Data Hiding

In this method the formation of QR Code from the encrypted message is discussed. The first step in creating a QR code is to create a string of data bits. This string includes the characters of the original message that you are encoding, as well as some information bits that will tell a QR decoder what type of QR Code it is.

After generating the aforementioned string of bits, we use it to generate the error correction code words fro the QR Code. QR Codes use Reed-Solomon Error Correction technique.
N.B.: In coding theory, Reed-Solomon codes (RS codes) are non-binary cyclic error correction codes invented by scientists Irving S. Reed and Gustave Solomon.

After the generation of bit-string and error correction code words, the resultant data is used to generate eight different QR Codes, each of which uses a different mask pattern. A mask pattern controls and changes the pixels to light or dark ones, according to a particular formula. The eight mask pattern formulas are defined in the QR Code specification, which is referred at the time of mask generation needed for the QR Code generation. Each of the eight QR codes is then given a penalty score that is based on rules defined in the QR specification. The purpose of this step is to make sure that the QR code doesn't contain patterns that might be difficult for a QR decoder to read, like large blocks of same-colored pixels, for example. After determining the best mask pattern, the QR Code which uses that mask pattern is generated and shown as an output.

If the size of the encrypted message becomes more than 1,264 characters then the characters appearing after 1,264 characters are used separately to generate another QR Code and the above mentioned process is repeated until and unless the total encrypted message is converted to QR Code(s).

#### a.     Binary String Generation

The first step of creating a QR code is to generate a binary string that includes the data and has information about the mode of encoding, as well as the length of the data. To generate the binary string, we first have to create the mode indicator, which will indicate the type of data (message) to be encrypted and this method is achieved by referring to a mode indicator table. In Fig 1.1 we can see the table fro mode-indicator.

| Bit string | Data mode |
|---|---|
| 0001 | Numeric Mode |
| 0010 | Alphanumeric Mode |
| 0100 | Binary Mode |
| 1000 | Japanese Mode |

Fig 1.1: Mode-indicator Table

For example, if the data, which is to be converted to QR Code, is 'I love you' then it is of Alphanumeric type and the mode-indicator for it is 0010.

After this step, the length of the total number of characters in the message (data) is calculated and then convert that into binary. For example, for the message 'I love you', the length of the string is 10, and binary of 10 is 1010. While encoding the length of data, the encoding is to be done using specific number of bits. The table for encoding the data using a specific number of bits is given in Fig 1.2.

**Versions 1 through 9**
- Numeric mode: 10 bits
- Alphanumeric mode: 9 bits
- Binary mode: 8 bits
- Japanese mode: 8 bits

**Versions 10 through 26**
- Numeric mode: 12 bits
- Alphanumeric mode: 11 bits
- Binary mode: 16
- Japanese mode: 10 bits

**Versions 27 through 40**
- Numeric mode: 14 bits
- Alphanumeric mode: 13 bits
- Binary mode: 16 bits
- Japanese mode: 12 bits

Fig 1.2: Table for Encoding Specific Number of bits

So, if we want to generate a Version 1 QR Code for the data (message) 'I love you', therefore the encoded length in binary format is: 000001010.

After this step, the encoded length is placed after the mode-indicator string to form the first level of binary string.

For the above example, the binary string so far becomes: 0010 000001010.

The Characters of the data are then paired up and the ASCII value of the first character of each pair is multiplied with 45 and the resultant is added to ASCII value of the second character of the pair. In this way the method is repeated for all the pairs and we convert the resultant of the pairs into binary format referring to the encoding table. If there is a single character left, which could not form a pair in the message (data), then only the ASCII value of that character is converted into binary format.
Now, that bit string is generated, it is to be made sure that it is of correct length. This depends on how many data bits are required to generate for the version and error correction that we are using.

b.  *Generating Error Correction*

As informed earlier, to generate the error correction code words **Reed-Solomon Code generation** technique is used. The first step to generating the error correction code words is to find out how many words you need to generate for your chosen QR version and error correction level. Reed-Solomon error correction uses polynomials. The first step is to construct a polynomial--called the **message polynomial**-- that uses the data blocks (binary strings) that was generated before. After this step, **generator polynomial** is created and polynomial division is performed on the generator polynomial to ultimately generate the error correction code words.

c.  *Generating Mask Pattern and QR Code*

After obtaining the error correction code words, the error code words are put after the bit string generated in the first step. Now that the data is encoded, all that is left to choose the best mask pattern. A mask pattern changes which bits are on and which bits are off, according to a particular rule. The mask patterns are defined in QR Code standard.

There are 8 mask patterns. Using mask patterns allows us to create eight different QR codes and then choose the one that will be easiest for a QR reader to scan. A particular QR code might have certain patterns or traits that make it difficult for QR readers to accurately scan the QR code.
After we generate the eight different QR codes internally, we give each one a penalty score according to the rules defined in the QR code standard. Then we output the QR code that has the best score.

D.  DECRYPTION

First the QR Code scanner scans the QR Code and generate the encrypted message and then decryption method is applied on that encrypted message. Decryption is actually the reverse process of the encryption. So basically, reverse engineering of all the steps of encryption (SD-EQR) are done and we get back the actual message, which was encrypted. Of course to get the actual message we need to provide the Symmetric Key, which was used in the first level of encryption.

E. RESULTS AND DISCUSSIONS

Encryption:

Here, the result of the above mentioned SD-EQR data hiding method is being showed with the result and limitations to this method is also being discussed.

Let us say, a person wants to send a message to another person and let the message be 'I love you ÿþý' and the password, which he use to encrypt the message is 'Hello World', then the following result is generated:

Message:  I love you ÿþý
Password: Hello World
Generated QR Code:

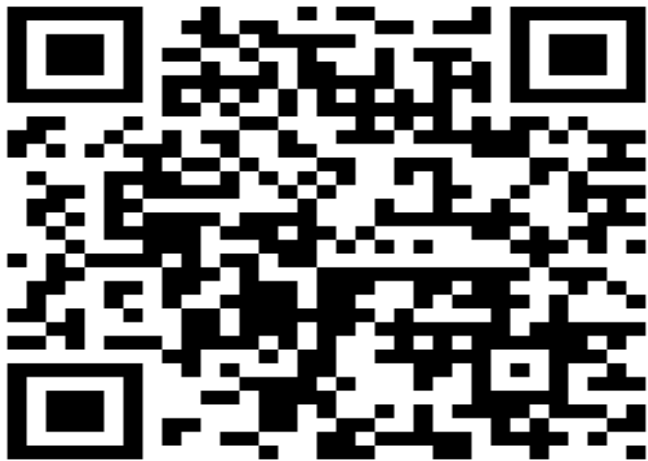

Message: I love you ÿþý

Limitations:

We have only applied this method for text messages and the results are not applied on any other type of message, i.e. on images, other small multimedia files, etc. But, the algorithm of SD-EQR holds a lot of potential and can be used for different types of messages.

i. Abbreviations and Acronyms

    a. Equations

The Equation used in Code Generation from the Symmetric Key provided by the user:

$N = \lim i \to 1 \; to \; plen \; \sum Di \times (plen)^2$,
N=Number generated from the password

**Code** $= \lim i \to 1 \; to \; len \; \sum ni,$    _(i)
$N = n_1 n_2 n_3 n_4 \ldots \ldots n_{(len)}$ and
$n_{(len)}$ = last digit of the number N

Addition of Code to Each Character and Unicode Generation:

**Value** = Code + ASCII($L_i$);    _(ii)
UniCode($L_i$) = '&#Value'; where i = 1 to n and $L_n$ = last letter of the last word of the message to be encrypted.  _(iii)

A. *Authors and Affiliations*

Somdip Dey (Author) is a student of B.Sc (Hons) Computer Science and affiliated to Department of Computer Science, St. Xavier's College [Autonomous], Kolkata, India.

B. *Identify the Headings*

  A. INTRODUCTION

  B. ENCRYPTION
- Use of Symmetric Key
- Generation of Code from Symmetric Key
- Addition of Code to Each Letter

*II. REVERSING THE ENCRYPTED MESSAGE*

*III. XOR-ING THE ENCRYPTED MESSAGE*

  C. QR CODE GENERATION AND DATA HIDING
- Binary String Generation
- Generating Error Correction
- Generating Mask Pattern And QR Code

  D. DECRYPTION

  E. RESULTS AND DISCUSSIONS
- Encryption
- Limitations

*i. Abbreviations and Acronyms*

*a. Equations*

*A. Authors And Affiliations*

*B. Identify the Headings*

  F. CONCLUSION

ACKNOWLEDGEMENT

REFERENCES

### F. CONCLUSION

In this SD-EQR method, only the encryption of text messages is shown. But, this method has a very large scope. Since Unicode format is used for encryption, this method can be used to encrypt any type of message or file (picture, video, audio, etc.) and send it to the receiver safely or the method can also be used to store important data or information safely. The inclusion of QR Code adds an extra level of security to the encrypted message and the receiver can access the original message very quickly, just by scanning the QR Code and decrypting it using a software, which uses the above mentioned SD-EQR algorithm. Even other encryption techniques / methods can be used with this method to add more level of security to the data or the message. Future scope of this technique is big and can also be implemented in daily life use, but, the use of this technique in reality depends and varies from user to user.

ACKNOWLEDGMENT

Somdip Dey expresses his gratitude to all his fellow students and faculty members of the Computer Science Department of St. Xavier's College [Autonomous], Kolkata, India, for their support and enthusiasm. He also thanks Dr. Asoke Nath, professor and founder of Computer Science Department of St. Xavier's College (Autonomous), Kolkata, for his constant support and helping out with the preparation of this paper.